\documentclass[aps,prb,twocolumn,showpacs,preprintnumbers,superscriptaddress,amsmath,amssymb,longbibliography]{revtex4-2}
\usepackage{graphicx}% Include figure files
\usepackage{subcaption}
\usepackage[normalem]{ulem}
\usepackage[svgnames]{xcolor}
\usepackage{dcolumn}% Align table columns on decimal point
\usepackage{bm}% bold math
\usepackage{amsfonts}
\usepackage{amsmath}
\usepackage{amssymb}
\usepackage{color}
\usepackage{enumerate}

\begin{document}

\title{Electronic and magnetic properties of (NdNiO$_3$)/(La$_{2/3}$Sr$_{1/3}$MnO$_3$) superlattices: a DFT+U perspective}% Force line breaks with \\

\author{Henrique M. M. Cardoso}
\affiliation{Departamento de Física, Universidade Federal de Minas Gerais, C.P. 702, 30123-970 Belo Horizonte, Minas Gerais, Brazil}
\author{Maria C. O. Aguiar}
\affiliation{Departamento de Física, Universidade Federal de Minas Gerais, C.P. 702, 30123-970 Belo Horizonte, Minas Gerais, Brazil}
\author{Cinthia Piamonteze}
\affiliation{Swiss Light Source, Paul Scherrer Institut, CH-5232 Villigen, Switzerland}
\author{Walber H. Brito}
\email{walber@fisica.ufmg.br}
\affiliation{Departamento de Física, Universidade Federal de Minas Gerais, C.P. 702, 30123-970 Belo Horizonte, Minas Gerais, Brazil}

\date{\today}

\begin{abstract}

The electronic and magnetic properties of NdNiO$_3$ (NNO) thin films are very sensitive to epitaxial strain and to the proximity to ferromagnetic oxides. In this work, we investigate the structural, electronic, and magnetic properties of NdNiO$_3$/La$_{2/3}$Sr$_{1/3}$MnO$_3$ (NNO/LSMO) superlattices under the epitaxial strain of NdGaO$_3$ (NGO) using density functional theory with Hubbard U correction (DFT+U) calculations.
Our findings reveal that LSMO induces insignificant structural distortions on the NiO$_6$ octahedra, and that there is a negligible charge transfer between LSMO and NNO.  More importantly, we find an intricate magnetic order regarding the interfacial Ni local moments, where we observe a coexistence of ferromagnetic interactions between interfacial Ni and Mn atoms, and antiferromagnetic interactions between inner Ni ions, which results in an overall ferromagnetic coupling across the interface.
We also observe the metallization of NNO through the emergence of a half-metallic density of states in the same spin channel as LSMO, which becomes suppressed the further we move away from the interface. 
Our theoretical findings agrees well with recent experimental data, shedding light on the complex interplay between the induced magnetic order and the electronic properties of NNO.

\end{abstract}

\maketitle

\section{\label{sec:int}Introduction}

The family of rare-earth nickelates, denoted as RENiO$_3$ (RE = La, Pr, Nd, Sm, ..., Lu), represents a distorted perovskite family that has garnered significant interest within the scientific community for many years. Due to the Ni-$3d$ derived low-energy electronic states, these materials exhibit properties heavily influenced by the strong electronic correlations and long-range magnetic ordering. As a result, the phase diagram of these materials exhibit a rich variety of phases characterized by an intricate interplay of electronic, magnetic, and structural degrees of freedom~ \cite{Torrance,Medarde_1997,G.Catalan}.
%%phases that are closely interconnected . 
From a theoretical standpoint, this complexity provides an excellent platform for studying the interactions and the effects of the different mechanisms in the physics of correlated electron materials. Furthermore, understanding the intricate interplay among spin, charge, and orbital degrees of freedom, and finding ways to manipulate these properties through confinement and proximity effects, holds great potential for exciting technological applications, particularly in the field of electronics~ \cite{J.Mannhart,Yang}.

Except for LaNiO$_3$, all bulk rare-earth nickelates exhibit a temperature dependent metal-insulator transition (MIT) in their phase diagrams  \cite{Medarde_1997,G.Catalan}. These MITs are attributed to what is known as a site-selective Mott transition  \cite{site_selective_mott,Johnston,Green,Haule2017-gc}. In this process, alongside the MIT, a structural phase transition occurs, changing the crystal structure from orthorhombic ($Pbnm$ space group) to monoclinic (P2$_1$/n)  \cite{Alonso_1,Alonso_2,Zaghrioui}. In this latter phase, a subsequent breathing distortion of the NiO$_6$ octahedra is observed. The Ni-O bond lengths alternate between large and small, producing a rock-salt-like pattern of large and small octahedra (see Appendix~\ref{NNO_bulk}). The Ni atoms within the large octahedra (referred to here as ``Ni$_{L}$") have longer Ni-O bond lengths, leading to reduced electron hopping and Mott localization. Conversely, the Ni atoms within the small octahedra (referred to here as ``Ni$_{S}$") heavily hybridize with surrounding oxygens, giving rise to a bonding-antibonding-like hybridization gap. The combined effect of these two phenomena results in an insulating state. At a specific N\'eel temperature ($T_N$), which can vary among different rare-earth nickelates. There is also an antiferromagnetic (AFM) ordering that follows a Bragg vector $\mathbf{q_0}$=(1/2,0,1/2)$_o$ in orthorhombic notation, which corresponds to $\mathbf{q_0}$=(1/4,1/4,1/4)$_{pc}$ in pseudocubic notation  \cite{Neutron-diffraction-1,Neutron-diffraction-2}.

As we vary the rare-earth element, the perovskite octahedra undergo rotations influenced by the atomic radii ratio of the involved elements  \cite{G.Catalan,Catalano_2018}. The degree of distortion can be quantified by measuring the deviation of the Ni-O-Ni bond angle from the ideal perovskite value of $180$ degrees. From an electronic perspective, tilting the Ni-O-Ni bond angle decreases hopping values, favoring electronic localization and an insulating state. In the case of Lutetium (Lu), the heaviest rare-earth element with the smallest atomic radius, these distortions are maximal, resulting in the highest temperature of the metal-insulator transition ($T_{MI}$). As the atomic radius of the rare-earth element increases, the distortions decrease, leading to a lower $T_{MI}$. For Neodymium (Nd), the $T_{MI}$ coincides with the ($T_N$) at 200 K, indicating a strong interplay between these two degrees of freedom.

One of the most recent studied methods for manipulating the metal-insulator transition (MIT) in these materials involves constructing heterostructures, superlattices, and thin films  \cite{Catalano_2018,Scherwitzl_2,Catalan_aleatorio,Middey}. These approaches offer additional control over key `knobs" such as strain, charge transfer and magnetic coupling. 
%The former can be affected by the so-called magnetic proximity effect. 
In this context, the interface between nickelates and mixed valence manganites comes to light. La$_{2/3}$Sr$_{1/3}$MnO$_{3}$ (LSMO) is a widely studied ferromagnetic (FM) half-metal, which has been observed to induce interesting effects on NNO when put in close proximity. Recent studies show emergent phenomenon such as novel interfacial ferromagnetic phases \cite{XMCD_LSAT_source}, which are usually attributed to charge transfer across the interface. Hole transfer from $Ni^{3+}$ to $Mn^{3+}$ would result in $Ni^{2+}$ and $Mn^{4+}$ at the interface, which, in turn, would favor the ferromagnetic $Ni^{2+}-O^{2-}-Mn^{4+}$ interaction. This charge transfer has, however, been shown to be strongly strain dependent \cite{strain_charge_transfer}. In fact, M. Caputo \textit{et al.}~ \cite{Caputo} reported a system of NNO/LSMO under NdNiO$_{3}$ (NGO)(110)$_o$ strain in which there was observed no significant charge transfer. It is important to mention that for NNO, the NGO strain is tensile  \cite{Catalano_NGO}, whereas for LSMO, it is compressive  \cite{lsmo_ngo}. Intriguingly, even in the absence of charge transfer, XMCD measurements revealed the emergence of a net magnetic moment in the Ni atoms, which are ferromagnetically coupled to the Mn atoms. In the absense of charge transfer, this effect was attributed to the stray field from the LSMO layer, which would align the spins at Ni sites. The authors also demonstrated, through ARPES and resistivity measurements, the suppression of the NNO's MIT due to the proximity to LSMO. 
These findings suggest that the electronic states of NNO are very sensitive to the emergent magnetic ordering induced by the proximity to LSMO. However, the precise mechanisms at play, their relation with the structural distortions, and the magnetic structure of these new phases are not well understood.

To gain a better understanding of the effects of the LSMO on the structural, magnetic, and electronic properties of NdNiO$_3$, we performed DFT+U calculations for NGO-strained superlattices composed of NNO and LSMO. Our findings indicate that there is no significant charge transfer between the Mn and Ni atoms, in good agreement with the XAS experiments reported in the supplementary material of  \cite{Caputo}. Moreover, we observe that the coupling at the interface is nontrivial, with both ferromagnetic and antiferromagnetic interactions occurring among different ions. Nevertheless, a net in-plane magnetic moment arises, primarily from the Ni$_{S}$ ions near the interface, which is not observed in NNO bulk. This net magnetic moment is ferromagnetically coupled with the interfacial Mn atoms, which is also in good agreement with the XMCD measurements of  \cite{Caputo}. 
Our calculations also reveal the presence of almost degenerated magnetic configurations. 
Finally, we observe the metallization of NNO, with semi-occupied Ni-3$d$ states concerning only one spin channel, resulting in a half-metallic ground state, in good agreement with the experimental observation of resolute metallicity in this interface by  \cite{Caputo}. 

The organization of our paper is as follows: In Sec. \ref{sec:mod}, we discuss the models and methods used in our theoretical analysis. We first present the structural models employed, followed by a discussion of the computational methods applied. In Sec. \ref{sec:res}, we present our results and discussions. We begin with an analysis of charge transfer and structural deformation, followed by a detailed examination of the magnetic configuration obtained from our calculations. This is followed by a study of the electronic structure, based on the electronic density of states of the superlattices, and we finish with a subsection discussing the effects of non-collinearity. Finally, in Sec. \ref{sec:conc}, we draw the conclusions of our paper.

\section{\label{sec:mod}Models and Methods}

\subsection{\label{subsec:struct_model} Structural models}

To model the interface between NNO and LSMO, we constructed a series of (001)$_{pc}$ NGO-strained superlattices (SL). The values of the NGO lattice parameters used in pseudocubic notation were $a_{pc} = 5.4176$~\AA{} and $b_{pc} = 5.4952$~\AA{}~ \cite{NGO_parameters}.

\begin{figure*}
\includegraphics[width=1.00\linewidth]{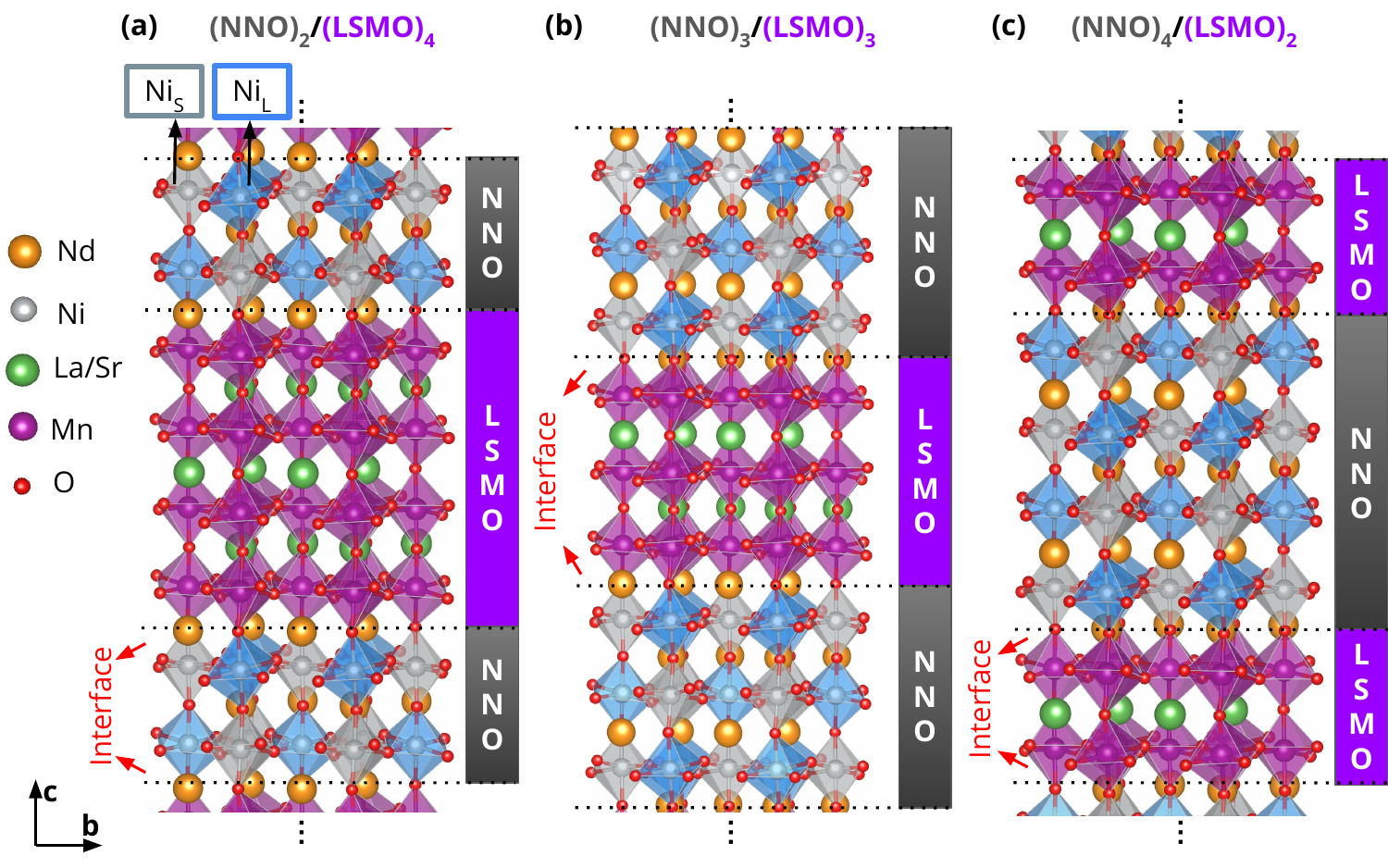}% Here is how to import EPS art
\caption{\label{modelos} Relaxed structures of (NNO/LSMO) superlattices  addressed in this work. The \textbf{c} and \textbf{b} axis are shown, and the interface is indicated by the horizontal dashed lines. The breathing distortions on NNO are shown as the blue and gray octahedra (small and large). In (a), (b), and (c) we show the obtained structures of (NNO)$_2$/(LSMO)$_4$, (NNO)$_3$/(LSMO)$_3$, and (NNO)$_4$/(LSMO)$_2$ superlattices. Images generated by the VESTA software~ \cite{VESTA}.}
\end{figure*}

It is usual to denote perovskite superlattices based on the number of ideal perovskite unit cells (u.c.) along the $c$-axis. 
%%Each ideal perovskite u.c is equivalent to an oxygen octahedron with a transition-metal ion inside. 
To investigate the possibility of different physics arising from differently arranged superlattices, we constructed three different structural models, varying the amount of NNO and LSMO unit cells along the $c$-axis. Within our notation, we considered superlattices featuring 2 NNO u.c. and 4 LSMO u.c., or concisely a (NNO)$_2$/(LSMO)$_4$ SL,  (NNO)$_3$/(LSMO)$_3$ SL, and (NNO)$_4$/(LSMO)$_2$ SL. These corresponding structural models are shown in Figures~\ref{modelos} (a), (b), and (c), respectively.

Additionally, it is important to mention that our superlattices extend significantly along the b-axis, as can be noticed in Figure~\ref{modelos}. When studying the bulk NNO, large $1\times 2 \times 2$ supercells are required to capture the bulk AFM Bragg vector $\mathbf{q_0}$ = (1/4, 1/4, 1/4)$_{pc}$ as well as the existence of the Ni$_L$(Ni$_S$) octahedra. We display the corresponding bulk configuration in Fig.~\ref{bulks}(b) of Appendix A. This represents the minimum structural configuration needed to accurately characterize the magnetic and structural properties of bulk NNO, as previously done in prior studies \cite{Haule2017-gc}.
When constructing our superlattice models, we did not impose the restriction that they should be able to precisely reproduce the exact magnetic ordering of the bulk, as the construction of the superlattice disrupts the symmetry along the c-axis. However, we argue that it is crucial to allow for the possibility of the bulk's magnetic ordering to be realized at least within the $a \times b$ plane.

\subsection{\label{subsec:comp_details} Computational Methods}

The calculations were carried out using density functional theory  \cite{dft}, as implemented in the Vienna ab initio simulation package (VASP)~ \cite{vasp_1, vasp_2, vasp_3}. For the post-processing of the VASP calculated data, we used VASPKIT~ \cite{vaspkit}, Sumo \cite{Sumo}, as well as the DensityTool code~ \cite{Densitytool}. Unless stated otherwise, our results are presented for spin arrangements treated at the collinear level, in accordance with previous theoretical works on rare-earth nickelates  \cite{NNO_DFTU_PARAMETERS, Haule2017-gc}.

To describe the effects of Sr doping in the La$_{2/3}$Sr$_{1/3}$MnO$_{3}$ material, we employed the Virtual Crystal Approximation (VCA)  \cite{VCA}, a method known to yield satisfactory results for LSMO~ \cite{lsmo_vca_1, lsmo_vca_2}. The choice to use the VCA stems from the considerable difficulty in constructing a realistic superlattice of doped LSMO. Doing so would require a substantially larger structure, or we would have to work with unrealistic configurations.

We represented the wave functions using the projector augmented wave (PAW) method  \cite{PAW}. A plane wave cutoff of 550 eV was set. For Nd, the additional 4f electrons were kept frozen in the core of the pseudopotentials. The PBEsol exchange-correlation functional was applied  \cite{PBEsol}. To take into account strong electronic correlations at a static mean-field level, we employed the DFT+U method  \cite{dft_u, lda_u}, following Liechtenstein \textit{et al.}'s approach~ \cite{Liechtenstein}. This method relies on the use of two parameters, one related to the on-site (local) coulomb repulsion energy, called U, and another associated with the Hund's coupling energy, called J. The parameters used for the Mn-3$d$ states were U = 5 eV and J = 1 eV  \cite{Vitao}. As for the Ni-3d states, the values of U = 2 eV and J = 0 eV were employed  \cite{NNO_DFTU_PARAMETERS}.

For the structure relaxation, we utilized a $6 \times 4 \times 2$ k-point grid and continued the calculations until the forces dropped below 0.005 eV/\AA. The supercell's shape was allowed to relax exclusively along the c-axis while maintaining the NGO strain constant. In the self-consistent calculations, we employed a k-point grid of $10 \times 8  \times 4$, with a convergence threshold set at $1\times10^{-6}$ eV.

\section{\label{sec:res}Results and Discussions}

\subsection{\label{subsec:chg_trans} Charge transfer and structural deformations}

The reorganization of electric charge occurring at the boundary between two different materials is something to be expected, as discrepancies in their electronic chemical potentials can induce the exchange of electric charge across this boundary. This charge exchange can modify the characteristics of the material, such as spin and orbital properties, leading to the emergence of new electronic and magnetic states within complex oxide heterostructures \cite{Middey,oxide-interfaces}. To evaluate if there is any charge transfer at the NNO/LSMO interface, we calculated the variation of the 3$d$ orbital occupation on the Ni and Mn atoms in comparison with the occupations of the bulk: $\Delta n$ = $n_{SL}$ - $n_{bulk}$. Here, we are interested solely in the effects of the interface, and not of the strain. The strain can create distortions which might change the values of the occupations by itself. To avoid such an interference with our results, we performed calculations for the bulk of the materials, imposing the same NGO strain on the $a \times b$ plane as the superlattices, while allowing them to relax exclusively in the c-axis direction. The 3$d$ occupancy of Ni and Mn atoms were obtained by integrating the 3$d$ projected density of states (PDOS) up to the Fermi energy. Overall, we find a negligible charge transfer between LSMO and NNO, with maximum $\Delta n$ of around 0.01 electrons in the case of Mn atoms in (NNO)$_4$/(LSMO)$_2$ SL.

Another important effect introduced by strain and confinement is the distortion of the lattice. 
In the perspective of a site-selective Mott insulator, one of the most important factors to  consider is the breathing distortion parameter $\delta d$. This parameter is used to represent the magnitude or amount of the octahedra distortion. It can be defined as the difference between the Larger/Smaller (L/S) bond lengths and the average bond length ($d_{0}$): $d_{L/S} = d_{0}\pm \delta d$. Previous theoretical studies have shown that the ground state of the rare-earth nickelates can be roughly understood as coexistent $d^8\underline{L^0}_{S=1}$ and $d^8\underline{L^2}_{S=0}$ states, where S denotes the spin, $d^8$ specifies the occupancy of the Ni d orbitals, and $\underline{L^2}$ or $\underline{L^0}$ represents the ligand holes on the oxygen \cite{Johnston,Green}. When considering the breathing distortion, the small octahedra collapse into the $d^8\underline{L^2}_{S=0}$ state, while the big octahedra collapse into the $d^8\underline{L^0}_{S=1}$ state. The bigger the breathing distortion parameter $\delta d$, the greater the effects on the magnetic and electronic properties of the material. 

In Table \ref{distortions}, we summarize the average breathing distortion parameter for the different analyzed structures. The idea was to understand how the strain and the interface would affect the breathing modes, and consequently, the physics of NNO. Surprisingly, we noticed no significant change in these parameters from one structure to the other. From the unstrained bulk to the NGO strained bulk, there is a slight increase of around 0.006 Å, which is insignificant. From the NGO strained bulk to the superlattices, the change is also very small, ranging from 0.001 Å for the (NNO)$_4$/(LSMO)$_2$ SL to 0.012 Å for the (NNO)$_2$/(LSMO)$_4$ SL. With that, we conclude that there is no significant change in the breathing distortion when applying the NGO strain or by the presence of the interface with LSMO. 

\begin{table}[h!]
\caption{\label{distortions} Distortions table. It summarizes the average distortion parameter ($\delta d$) and the average Ni-O-Ni bond angle ($\theta$) for different analyzed structures. }
\begin{ruledtabular}
\begin{tabular}{cccccccc}
 Structure & $\langle \delta d \rangle$(\AA)& $ \theta \langle$Ni-O-Ni$\rangle$($^{\circ}$) & $\langle$Ni-O-Mn$\rangle$($^{\circ}$) \\
\hline
Unstrained Bulk& 0.031 & 157.53 & - \\
NGO strained Bulk & 0.037 & 155.92 & -  \\

(NNO)$_2$/(LSMO)$_4$ SL & 0.025 & 156.00 & 154.96 \\
(NNO)$_3$/(LSMO)$_3$ SL  & 0.033 & 155.89 &155.20  \\
(NNO)$_4$/(LSMO)$_2$ SL & 0.036 & 155.74 & 155.26 \\

\end{tabular}
\end{ruledtabular}
\end{table}

There is also another important structural factor to consider in our distortion analysis. For the rare-earth nickelates in particular, there is a strong connection between the Ni-O-Ni bond angle and the metal-insulator transition temperature - $T_{MI}$  \cite{G.Catalan,Catalano_2018}. As we decrease the size of the rare-earth radius, the Ni-O-Ni bond angle decreases, and the $T_{MI}$ increases. In Table \ref{distortions}, we show the average Ni-O-Ni bond angle ($\theta$) for the different analyzed structures. The known $\theta$ value for NNO is approximately $157^\circ$  \cite{Medarde_1997}, which is in good agreement with our value of $157.53^\circ$. When we apply the NGO strain, which is tensile, the effect is to slightly decrease the bond angle to $155.92^\circ$. 

When constructing the interface, there is also no significant change in the overall Ni-O-Ni bond angle. When compared to the NGO strained bulk, the changes range between $0.03^\circ$ for the (NNO)$_3$/(LSMO)$_3$ SL, to $0.18^\circ$ for the (NNO)$_4$/(LSMO)$_2$ SL, which amounts to an almost insignificant variation. On Table \ref{distortions}, we also show the average Ni-O-Mn bond angle across the interface for the different superlattice models. Although there seems to be an overall tendency for them to be slightly smaller than the average Ni-O-Ni angle, the change is also very small, ranging from $-1.04^\circ$ for the (NNO)$_2$/(LSMO)$_4$ SL, to $-0.48^\circ$ for the (NNO)$_4$/(LSMO)$_2$ SL, giving another strong indication that the interface does not introduce relevant lattice distortions on the system.

\subsection{\label{subsec:mag_conf} Magnetic configuration}

One of the most intriguing questions raised by Caputo and coworkers in Ref. \cite{Caputo}  was how the magnetic coupling at the NNO/LSMO interface actually occurs. To evaluate the detailed magnetic coupling at the interface, we calculated the spin density of our superlattice models. Spin density is defined as the difference between the spin-up and spin-down charge densities, providing a clear depiction of the material's magnetic configuration.

\begin{figure*}
\includegraphics[width=1.00\linewidth]{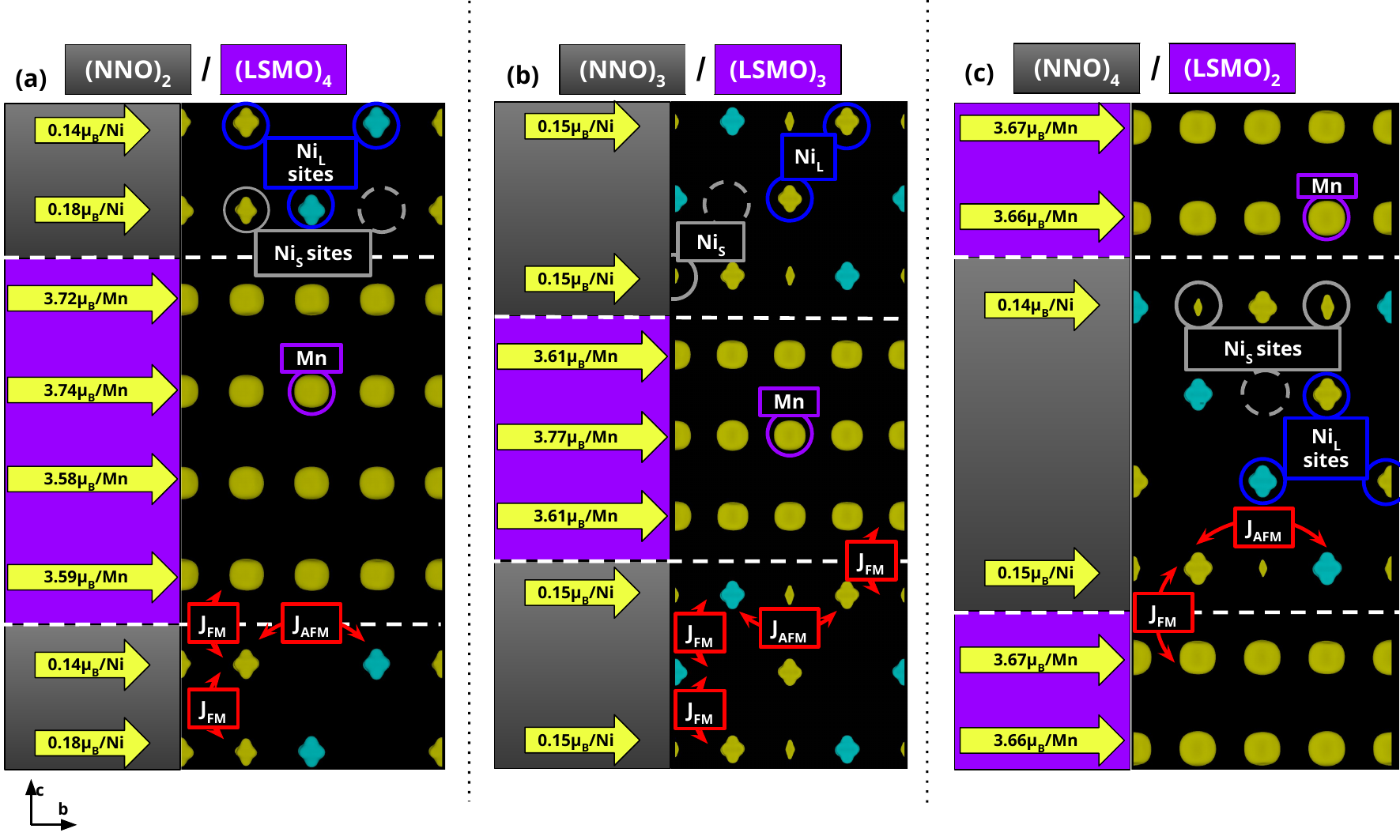}% Here is how to import EPS art
\caption{\label{SPIN_DENSITY} Calculated spin densities for (a) (NNO)$_2$/(LSMO)$_4$, (b) (NNO)$_3$/(LSMO)$_3$, and (c) (NNO)$_4$/(LSMO)$_2$ superlattices. The blue isosurface indicates the spin-down density, while the yellow regions indicate the spin-up densities. Regions in black have negligible spin charge differences. The gray circles denote Ni$_S$ sites, with the solid-lined circles indicating the sites with a considerable magnetic moment, and the dashed-lined circles indicating those with negligible magnetic moment. The blue circles denote Ni$_L$ sites, and the purple circles denote Mn sites. The red boxes with arrows describe important magnetic exchange interactions between ions. On the left of each spin density plot, we display a schematic drawing of the materials and the interface, with purple blocks representing LSMO, and gray blocks represent NNO. The yellow arrows indicate the direction and value of the averaged magnetic moment on each plane.}
\end{figure*}

Figures~\ref{SPIN_DENSITY} (a), (b), and (c) show the calculated spin densities for (NNO)$_2$/(LSMO)$_4$, (NNO)$_3$/(LSMO)$_3$, and (NNO)$_4$/(LSMO)$_2$, respectively. The blue isosurface indicates the spin-down density, while the yellow regions indicate the spin-up densities. Regions in black have negligible spin charge differences. The gray circles denote Ni${_S}$ sites, the blue circles denote Ni${_L}$ sites, and finally, the purple circles denote Mn sites. On the left of the spin density plots, there is a schematic drawing that helps visualize the separation of the materials and the interface. The purple blocks represent the LSMO, while the gray blocks represent the NNO. Inside the schematic drawings, we display yellow arrows that indicate the direction of the net magnetic moment on each plane. For the planes with no arrows, the magnetic moment is close to zero. Within the arrows, we present the average magnetic moment per transition metal in that plane. As a comparison, the calculated bulk values of the average magnetic moment per ion per plane for NNO are zero. For LSMO it is 3.70 $\mu_{B}/Mn$.

Starting from the (NNO)$_2$/(LSMO)$_4$ model shown in Figure \ref{SPIN_DENSITY} (a), we can clearly observe a net magnetic moment per ion of approximately 0.14 and 0.18 $\mu_{B}/Ni$ on the interfacial NiO$_2$ planes. Furthermore, this net magnetic moment is ferromagnetically coupled with its neighboring MnO$_2$ planes, as indicated by the direction of the arrows. This result is in good agreement with the XMCD data reported on Ref. \cite{Caputo}.

Upon closer inspection of the magnetic configuration through the spin density plot on the right of Figure \ref{SPIN_DENSITY} (a), we find that, despite having this net ferromagnetic moment, the interfacial NiO$_2$ planes are not entirely ferromagnetic. Instead, the Ni${_L}$ sites inside the NiO$_2$ plane are still antiferromagnetically coupled with each other along the b-axis, similar to the bulk (see Figure \ref{bulks} (b)). On the other hand, some of the Ni${_S}$ sites at the interface have acquired a local magnetic moment not observed in our bulk calculations, as is highlighted by the solid-lined circles of Figure \ref{SPIN_DENSITY}. The appearance of this new magnetic moment appears to be a key insight from our calculations, as it is ferromagnetically coupled to the neighboring MnO$_2$ planes and constitutes the majority of the net ferromagnetic moment at the interface.

This result is particularly interesting and highlights that the magnetic configuration of this system is far from trivial, consisting of different magnetic couplings arising from the competition between multiple interactions.

\begin{figure}
\includegraphics[width=1.00\linewidth]{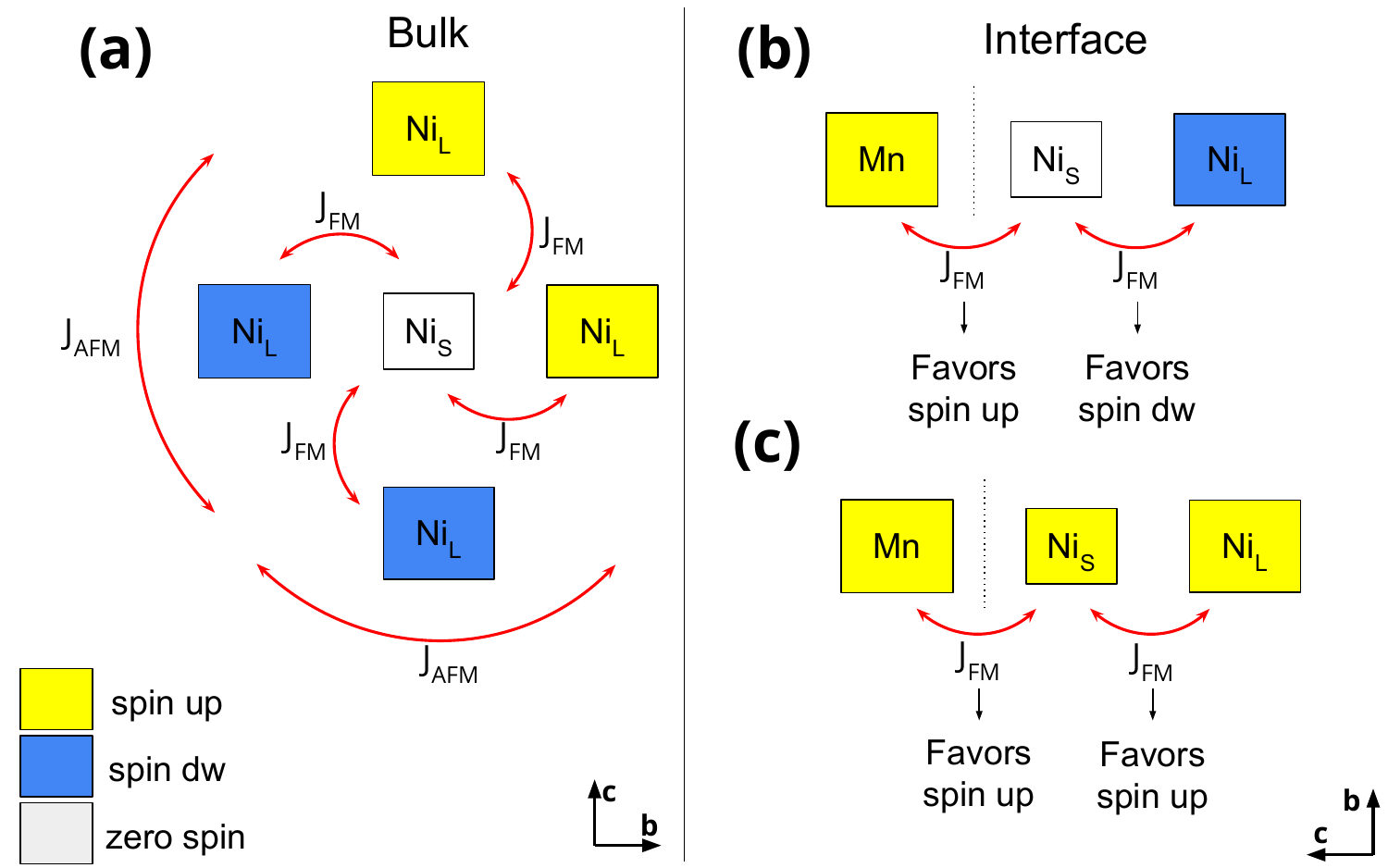}% Here is how to import EPS art
\caption{Simplified model of important magnetic exchange interactions between ions on the Bulk NNO (a), and on the interface NNO/LSMO (b), and (c). Ni$_L$ and Ni$_S$ indicate the large and small Ni sites, respectively, and Mn indicate the manganese. The yellow, blue and white boxes indicate a net spin up, down, and zero, respectively. \label{explicacao_spin} 
}
\end{figure}

In figure \ref{explicacao_spin} (a) we display a simplified model that attempts to describe the most important magnetic interactions of the bulk NNO on the c$\times$b plane. There is a (Ni$_L$)-(Ni$_L$) antiferromagnetic interaction between the Ni$_L$ sites, and a (Ni$_L$)-(Ni$_S$) ferromagnetic interaction between neighboring Ni$_L$ and Ni$_S$ sites. Given that each Ni$_S$ ion has the same number of spin-up and spin-down Ni$_L$ first neighbors, the spin-up and spin-down influence evens out, resulting in a zero net magnetic moment.

When we introduce an interface with a ferromagnetic system, this symmetry can be broken. In Figure \ref{explicacao_spin} (b) and (c), we show the two possible outcomes when we consider a (Mn)-(Ni) ferromagnetic interaction (J$_{FM}$) across the interface. In (b), we have a Ni$_S$ ion between a spin-up Mn and a spin-down Ni$_L$. In this situation, we have something similar to the bulk, where the spin-up and spin-down influences even out, and the Ni$_S$ acquires a negligible net spin. The other possible outcome is depicted on (c). In this case, the Ni$_S$ ion is between a spin-up Mn and a spin-up Ni$_L$. With no spin-down influence, the Ni$_S$ acquires a novel magnetic moment not present in the bulk. These two situations can be observed in the spin density plot of Figure \ref{SPIN_DENSITY} (a), where we circled in gray the two Ni$_S$ sites, with the solid-lined circles indicating the sites with a considerable magnetic moment, and the dashed-lined circles indicating those with negligible magnetic moment. The one on the right, with a dashed-lined circle, shows a spin density close to zero, depicting the (b) case of Figure \ref{explicacao_spin}. The one on the left, with a solid-lined circle, has a significant spin-up density, and it is described by Figure \ref{explicacao_spin} (c). 

Figure~\ref{SPIN_DENSITY} (b) shows our results for the (NNO)$_3$/(LSMO)$_3$ superlattice. The interesting addition made by this model comes from the fact that now we have a NiO$_2$ plane that is not located at the interface. Upon close inspection, we observed that, although the appearance of the novel magnetic moments on the interfacial Ni$_S$ ions is still present in this model (solid-lined gray circle on the bottom-left), the changes in the magnetic structure are exclusively located at the interface. The middle NiO$_2$ plane does not acquire a significant new magnetic moment (dashed-lined gray circle on the top-right), and, consequently, the net magnetic moment of the middle plane vanishes, as it is in the bulk. This is depicted by the absence of arrows related to the middle plane on the schematic image on the left of Figure \ref{SPIN_DENSITY} (b).

The lack of a novel magnetic moment can be understood again by the scheme in Figure \ref{explicacao_spin} (a). The middle Ni$_S$ ions are located at the center of four Ni$_L$, two with spin-up, and two with spin-down. Consequently, the influence of spin-up and down amounts to no spin favoring at all, resulting in a negligible spin for the ion.

Finally, in Figure \ref{SPIN_DENSITY} (c) we show the obtained spin-density for the (NNO)$_4$/(LSMO)$_2$ superlattice. This model reaffirms that the novel magnetic moment arising from the Ni$_S$ sites does not extend beyond the interface. What is also interesting to note, both in this model and in the (NNO)$_3$/(LSMO)$_3$ model, is that now, all the Ni$_S$ sites at the interface acquire a novel magnetic moment, even those followed on the c-axis by a spin-down Ni$_L$ site (left solid-lined gray circles on (b) and (c)). This did not happen in the (NNO)$_2$/(LSMO)$_4$ model and cannot be directly explained by the simplified model in Figure \ref{explicacao_spin}. We believe that the reason for this lies in the increased size of NNO in models (b) and (c), where long-range interactions start to play an important role.

That being said, the mechanism described by Figure \ref{explicacao_spin} still has a significant influence here. Even though the Ni$_S$ ions followed by a spin-down Ni$_L$ site do not have a zero net magnetic moment, they have a noticeably smaller magnetic moment than the ones followed by a spin-up Ni$_L$ site. From this, we draw two main conclusions. First, the appearance of novel magnetic moments at the interface will be universal for all models and is related to the coupling between the Mn-Ni ions across the interface. Second, the Ni$_S$ sites followed by a spin-down Ni$_L$ ion will have a smaller magnetic moment, but this discrepancy becomes less significant as we increase the size of the NNO film.

\subsection{\label{subsec:Metastable} Metastable states}

The results discussed so far were all obtained by relaxing the crystalline and magnetic structure from a starting point based on the bulk configurations. They were also the most stable configurations that we obtained. However, it is important to draw attention to the fact that there are different crystalline and magnetic configurations possible for this system. As reported in previous theoretical studies, the NNO system has a very sensitive magnetic configuration and is known to display different metastable states from a theoretical perspective  \cite{site_selective_mott,Catalano_2018}.

\begin{figure*}
\includegraphics[width=1.00\linewidth]{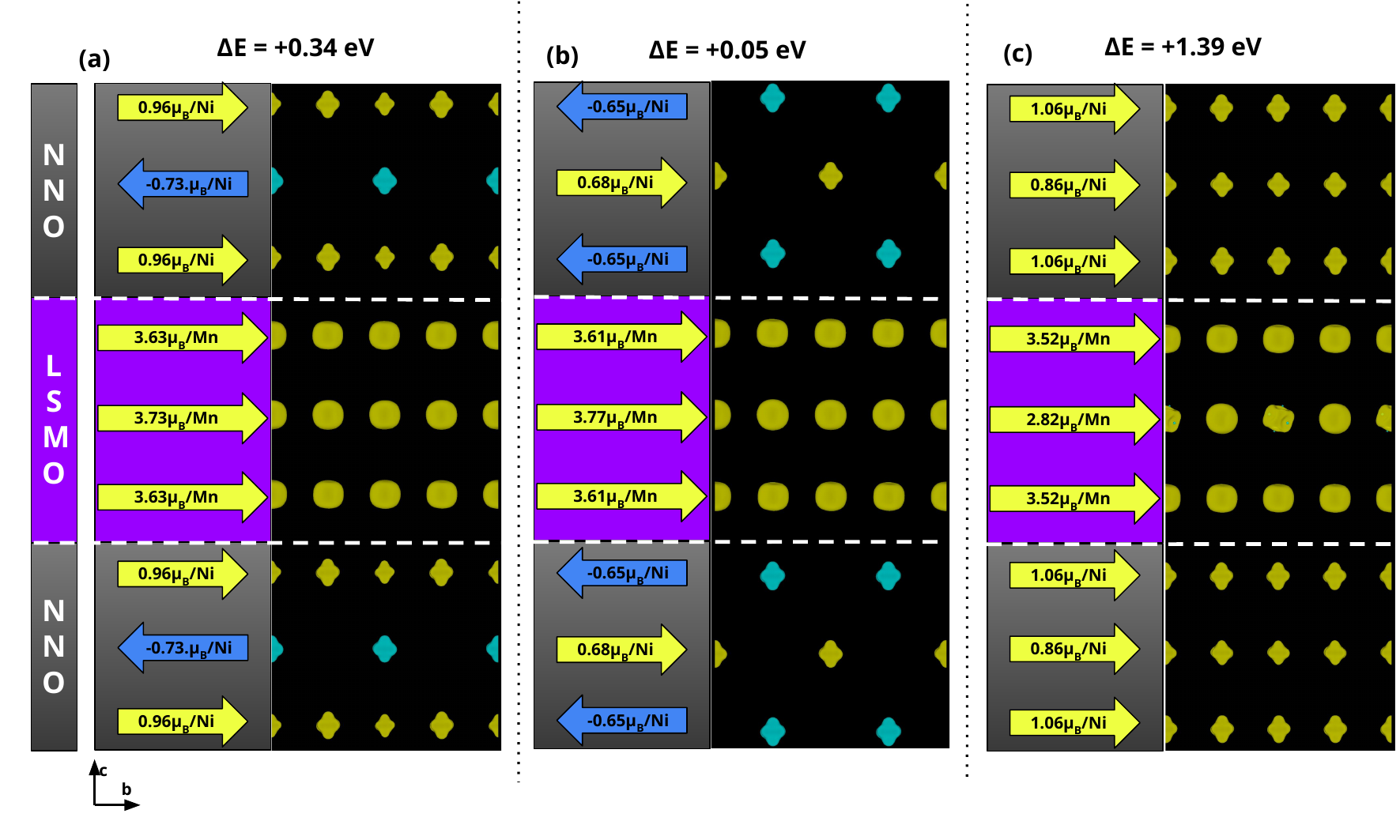}% Here is how to import EPS art
\caption{Spin densities for alternative magnetic configurations of the (NNO)$_3$/(LSMO)$_3$ model. On top of each image, the energy difference between them and the most stable state (Figure \ref{SPIN_DENSITY} (b)) is displayed. Colors and other notations are similar to the ones in Fig. 2. \label{alternative_spin}  
}
\end{figure*}

Figure~\ref{alternative_spin} shows three alternative magnetic configurations obtained by our calculations for the (NNO)$_3$/(LSMO)$_3$ superlattice. On top of each configuration, we present the energy difference $\Delta E$ when compared with the most stable result. If $\Delta E$ is positive, it means that this configuration is less stable (has greater total energy) than the one in our main result in Figure \ref{SPIN_DENSITY} (b). Our experience with this system's sensitivity to initial parameters leads us to believe that there are likely many other metastable states, but the three displayed here are noteworthy for a brief discussion.

The first two display ferromagnetic planes that are antiferromagnetically coupled to each other. In Fig.~\ref{alternative_spin} (a), we have ferromagnetic coupling at the interface, while in (b), we have antiferromagnetic coupling at the interface. What is most interesting about these two configurations is the fact that the energy difference between them and the main result is very small (smaller than 1 eV). This draws attention to the possibility of widely different states with negligible energy differences and sheds light on the difficulty of studying this system from a theoretical perspective. The configuration showed on Figure \ref{alternative_spin} (a) has a higher energy difference, making it less likely, but the net magnetic moment coming from the Ni atoms is approximately $0.40\mu_{B}/Ni$.
On (b) however, although the energy difference is very small, the net magnetic moment is approximately $-0.20\mu_{B}/Ni$, which means the overall Ni magnetic moment is aligned antiferromagneticaly with the Mn, which is not observed experimentally~\cite{Caputo}.

The last configuration, shown in Figure \ref{alternative_spin} (c), depicts an entirely ferromagnetic system, resulting in a net magnetic moment from the Ni atoms of approximately $0.99 \mu_{B}/Ni$, which is considerably larger than what is observed experimentally. Along with the large energy difference of $\Delta E = +1.39 \text{ eV}$, this makes it the least likely configuration we found. It is interesting to note that previous theoretical analyses by  \cite{Caputo} used a completely ferromagnetic NNO system to explain the novel metallicity observed experimentally at the NNO/LSMO heterostructure. Here, we demonstrate that while an entirely ferromagnetic system is possible, it is extremely unlikely due to its significantly higher energy compared to the most stable configuration we identified, as well as several other metastable states with lower energy. In the next section, we will show that this magnetic structure, although resulting in a metallic system, is not necessary for the metallization of NNO.

\subsection{\label{subsec:dos_analysis} Electronic density of states}

As mentioned in Section~\ref{sec:int}, one of the key findings reported on Ref. \cite{Caputo} was that the thin film of NGO-strained NNO/LSMO suppressed the NNO's metal-insulator transition. To assess whether our superlattice model captures this observed phenomenon, we analyzed the electronic properties of our system by examining its density of states (DOS).

\begin{figure*}
\includegraphics[width=1.00\linewidth]{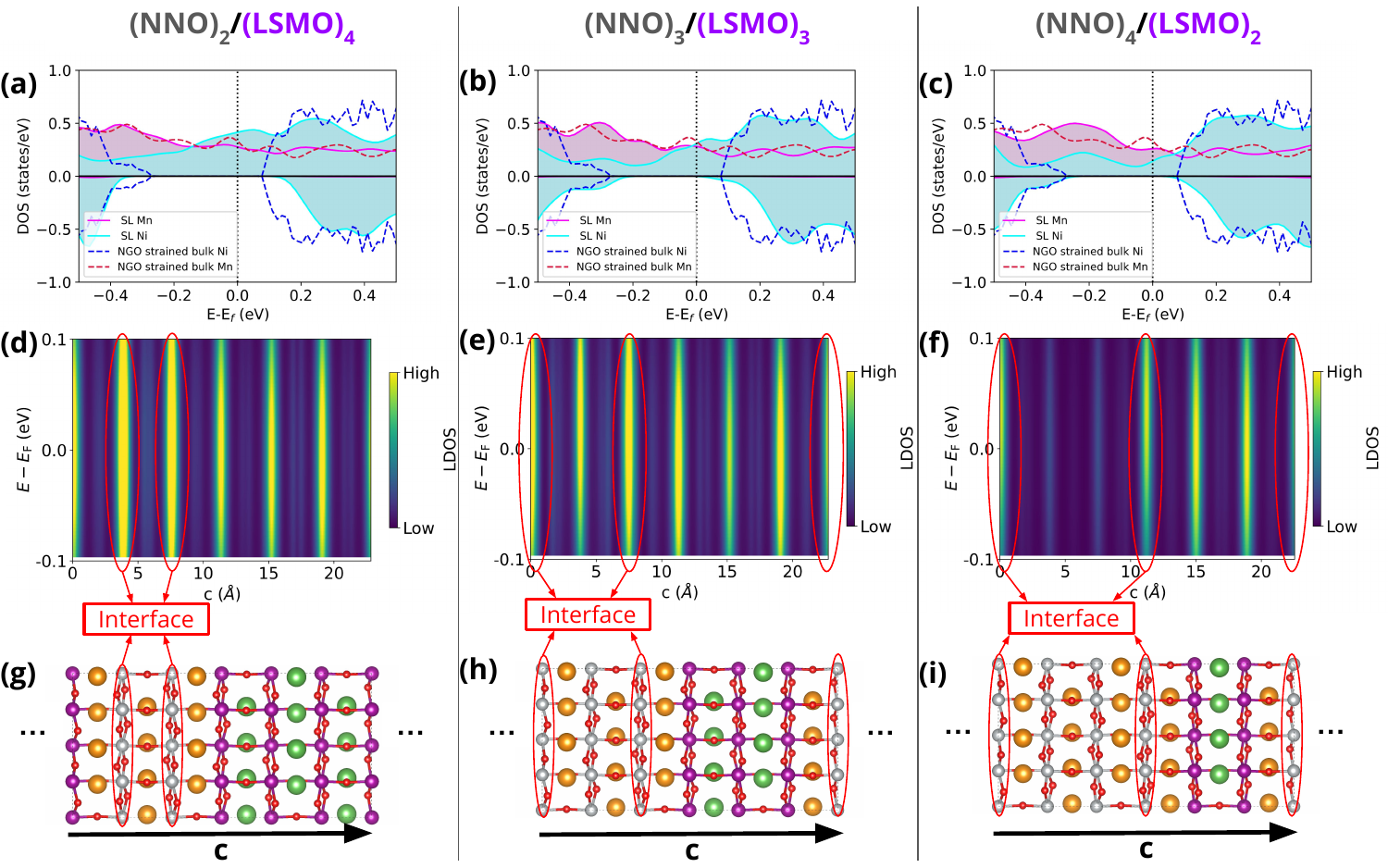}% Here is how to import EPS art
\caption{The Figure is divided into three panels: left, middle, and right. The left panel ((a), (d), and (g)) is related to the (NNO)$_2$/(LSMO)$_4$ SL, the middle panel ((b), (e), and (h)) is related to the (NNO)$_3$/(LSMO)$_3$ SL, and the right panel ((c), (f), and (i)) to the (NNO)$_4$/(LSMO)$_2$ SL. The top images, (a), (b), and (c) display the d-orbital projected density of states for each model, with the zero set at the Fermi energy. Positive and negative values of DOS indicate spin up and spin down, respectively. The dashed blue and red lines show the PDOS for the Ni and Mn in the NGO strained bulks. The cyan and magenta solid lines show the PDOS for Ni and Mn in the superlattices. The center images, (d), (e), and (f) display the LDOS along the c-axis of each model. The yellow and blue regions indicate, respectively, strong and weak density of states intensity. The bottom images (g), (h), and (i) contain the crystalline structure of each model, for better comparison with the LDOS plots. Both on the LDOS and the crystal structure images, the interface NiO$_2$ planes are circled.\label{dos_ldos}}
\end{figure*}

Figures~\ref{dos_ldos} (a), (b), and (c) show the plots of the obtained density of states for the three different models analyzed in our work. 
The dashed blue line represents the Ni's d orbital projected density of states for the NGO-strained NNO bulk, and the red dashed line represents the same for Mn in the LSMO bulk. These lines serve as parameters to evaluate how the density of states changes from the bulk to the superlattice. The magenta and cyan curves describe, respectively, the d orbital projected density of states of Mn and Ni in the superlattices.

As can be clearly seen by the dashed lines, the bulk LSMO is a half metal (having density of states crossing the Fermi energy only for spin-up electrons), while the bulk NNO is an antiferromagnetic (AFM) insulator, with symmetric spin-up and down density of states. 

Comparing the Mn density of states when transitioning from the bulk to the superlattice, we do not observe a significant change on all three models. Manganese continues to exhibit half-metal behavior on the superlattices, with no observable tendency to decrease or increase the density of states around the Fermi energy. This indicates that the interface does not significantly affect the Mn electronic structure.

On the other hand, comparing the Ni's density of states when transitioning from the bulk to the superlattice, we can clearly see that the insulating gap closes, giving rise to a half-metallic state with the same spin channel as the Mn. The presence of a finite density of states crossing the Fermi level indicates a metallic system, which is not observed in the bulk material and is a product of the interface between these two complex oxides. This result is not only intriguing, but it is also in agreement with the experimental findings reported by Caputo \textit{et al.}~ \cite{Caputo}.

We are also interested in the depth behavior of this emergent metallic state in the NNO. With that in mind, we analyzed the local density of states (LDOS) integrated over the $a \times b$ plane. This analysis allows us to evaluate the weight of the density of states as a function of the position on the c-axis, revealing how it changes as we move closer or further away from the interface. The relevance of analyzing differently sized superlattices (SLs) comes to light in this context.

Figures~\ref{dos_ldos} (d), (e), and (f) show the LDOS plots for the different models analyzed. On the y-axis, we have the energy in eV. The x-axis represents the c-direction (\AA) in real space. The color map shows the intensity of the density of states, where yellow indicates high intensity and dark blue indicates low intensity. The LDOS exhibits multiple very intense stripes regularly distributed along the c-direction. These stripes correspond to the MnO$_2$ or NiO$_2$ planes. As most of the states around the Fermi energy come from these planes, the LDOS intensity is distinctly stronger at their positions.

For better visualization, Figure \ref{dos_ldos} (g), (h), and (i) display the crystalline structure of each model, with the planes aligned with their respective stripes. This serves as a guide to better understand the meaning of the LDOS plots. To enhance clarity, the interfacial NiO$_2$ planes are circled in red on both the LDOS plots and the crystalline structure.

On the left panel of Figure \ref{dos_ldos}, we present our results for the (NNO)$_2$/(LSMO)$_4$ superlattice. All the Ni atoms in this particular model are located at the interface. From (d), we can see that the two NiO$_2$ planes show a very intense LDOS signal, indicating a robust metallic state, with numerous states crossing the Fermi level.

In the middle panel, we depict the (NNO)$_3$/(LSMO)$_3$ superlattice. This model now has two NiO$_2$ planes at the interface and one NiO$_2$ plane at the center. Our results also display a very strong signal coming from the interface for this model. Furthermore, the metallic state seems to extend beyond the interface between the two materials, as the middle NiO$_2$ plane also exhibits a strong LDOS, although slightly smaller than at the interface.

Finally, on the right panel, we analyze the (NNO)$_4$/(LSMO)$_2$ superlattice. This final model has four NiO$_2$ planes, two at the interface and two at the center. Our results clearly show that while the interface NiO$_2$ planes display a strong LDOS signal, the center ones do not. They still show a slim signal, but it has greatly diminished when compared to the interface, which is a noteworthy result. This indicates that the further away from the interface we get, the weaker the LDOS signal around the Fermi energy. We believe that, with larger models, far away from the interface the signal would eventually diminish to zero, and the insulator character from the bulk would be restored. However, reproducing this is impractical at the moment, given the computational cost of further increasing the model size.

As mentioned earlier, the intricate magnetic structure in this system appears to be limited to the interface, as the middle NiO$_2$ planes have not shown the appearance of significant new magnetic moments on the Ni$_{S}$ sites. However, the depth of the metallic state seems to extend beyond the interface planes, as shown in Figure \ref{dos_ldos} (e) and (f). Even though the density of states crossing the Fermi level decreases for the middle NiO$_2$ planes, they still exist, indicating that while this novel metallic state might be strengthened by the new magnetic ordering, it is not necessarily coupled to it. 

\subsection{Non-colinear Calculations}

\begin{figure*}
\includegraphics[scale=0.6]{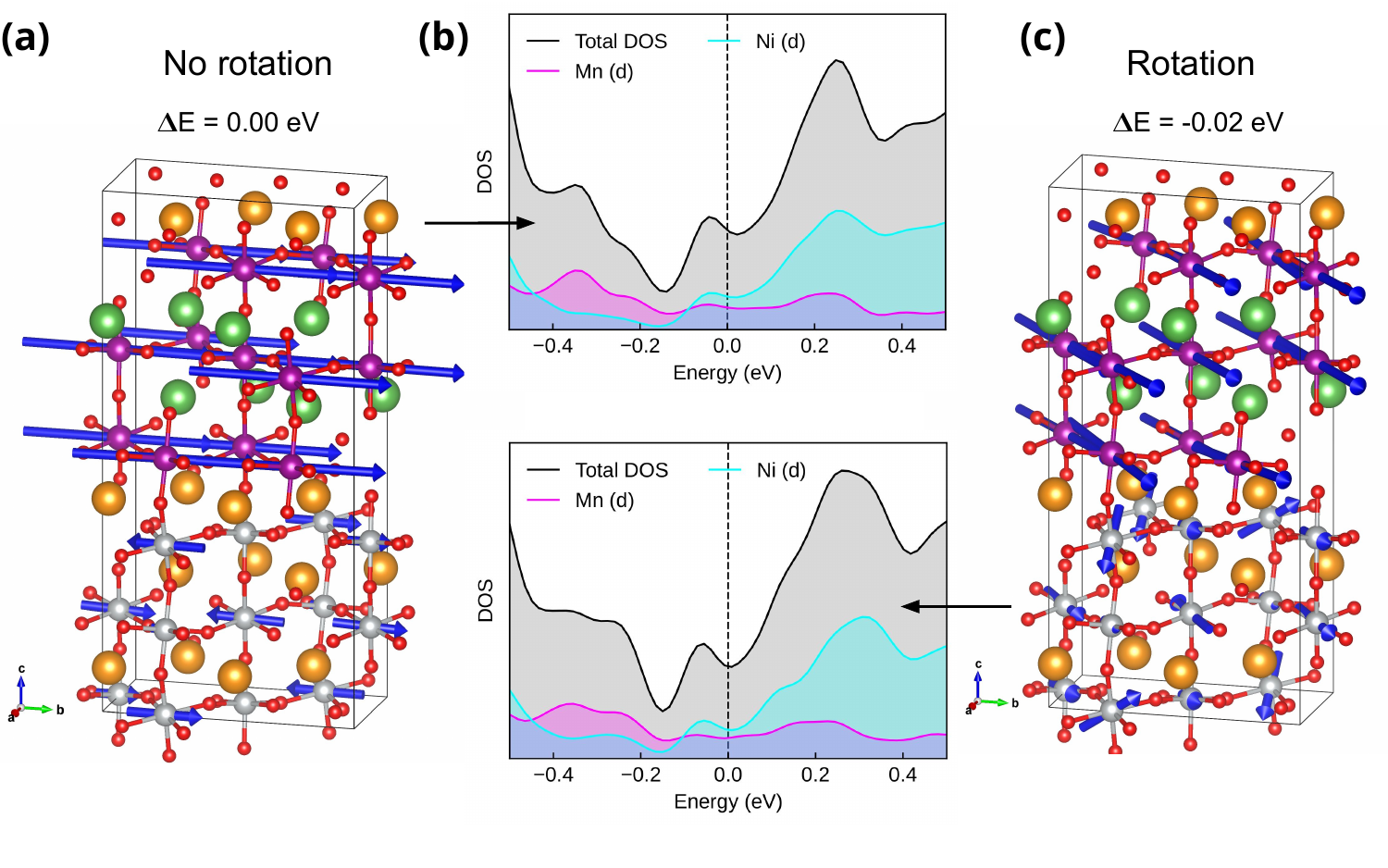}% Here is how to import EPS art
\caption{Spin-orbit non-collinear calculations. The blue arrows indicate the direction of the local magnetic moments. (a) Magnetic configuration with insignificant rotation of local magnetic moments. (b) Density of states for the two possible magnetic configurations displayed. (c) Magnetic configuration with significant non-collinearity between local magnetic moments.}\label{non_col_spin}
\end{figure*}

Previous reports on nickelate/manganite interfaces have shown that significant non-collinear effects can arise at these interfaces. Hoffman and co-workers \cite{Non-collinear_interface} observed oscillatory non-collinear magnetism in LaNiO$_{3}$/La$_{2/3}$Sr$_{1/3}$MnO$_3$ superlattices. Considering this, we performed non-collinear spin-orbit calculations for the (NNO)$_3$/(LSMO)$_3$ model to check for a different magnetic ground state. Figures \ref{non_col_spin} (a) and (c) show two different non-collinear magnetic orderings obtained from our calculations. The arrows indicate the direction of the local moments. In (a), we have a configuration with negligible rotation, while in (c), we display a different configuration with considerable non-collinearity. In (b), we show the electronic density of states for the configuration with no rotation (top) and the configuration with rotation (bottom). The two obtained configurations have  similar total energies, with the one including rotation being slightly more stable, with $\Delta E = -0.02$ eV. This confirms previous discussions that this system has many magnetic metastable states. Regardless, the spin-orbit non-collinear calculations do not significantly alter the effects observed in the collinear calculations. We observe that the ferromagnetic LSMO continues to induce small magnetic moments on the interfacial Ni${S}$ sites, which are coupled ferromagnetically (pointing in the same direction). Moreover, the density of states for both the rotated and non-rotated configurations present similar features indicating the metallization of NNO. This indicates that while there may be some non-collinear effects in this system, they do not affect our main findings obtained within the collinear approximations.

\section{\label{sec:conc}Conclusions}

In conclusion, our results demonstrate that the interface induces minimal structural distortions and insignificant charge transfer. Furthermore, we showed that the magnetic configuration of the interface is far from trivial and involves competing ferromagnetic and antiferromagnetic exchange interactions between the Ni$_{S}$, Ni$_{L}$, and the Mn atoms. There is, however, the appearance of a net magnetic moment in the interfacial Ni ions, which couple ferromagnetically with the Mn ions of LSMO.

We also observed the emergence of a half-metal density of states with the same spin channel as LSMO, originating from the d orbitals of the Ni ions. This metallic behavior is suppressed the further from the interface we go, suggesting that for a large enough model, an insulator regime would be reestablished far away from the interface. 

Our theoretical results are in good agreement with previously reported experimental data \cite{Caputo}, and help to shed light on the complexity and underlying physics taking place in this rather intricate system, representing a significant step in the study of these complex magnetic interfaces.

\begin{acknowledgments}

The authors acknowledge the financial support from the Brazilian agencies CNPq (in particular Grants 402919/2021-1 and INCT-IQ 465469/2014-0), CAPES, FAPEMIG. The authors also acknowledge the National Laboratory for Scientific Computing (LNCC/MCTI, Brazil) for providing HPC resources of the SDumont supercomputer (URL: http://sdumont.lncc.br), as well as the `Centro Nacional de Processamento de Alto Desempenho em São Paulo (CENAPAD-SP)", both of which have contributed to the research results reported within this paper. 

\end{acknowledgments}

\appendix

\section{NNO and LSMO bulk}\label{NNO_bulk}

\begin{figure}[h]!
\includegraphics[width=1.00\linewidth]{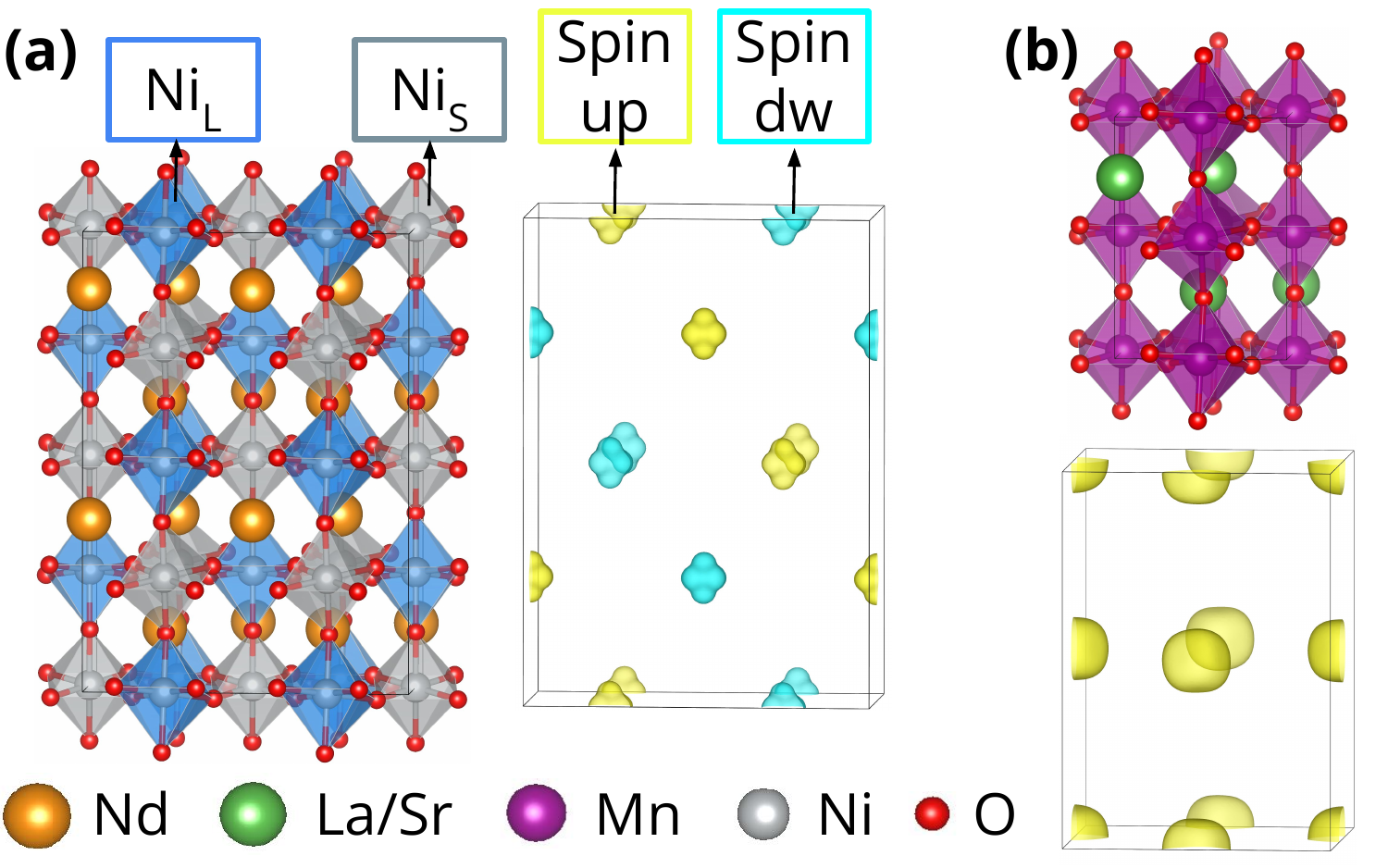}% Here is how to import EPS art
\caption{\label{bulks} (a) On the left, the bulk structure of NNO is shown. The octahedra have two different colors, as a consequence of the breathing distortions present on this material. The blue color is referring to the expanded octahedra (Ni$_{L}$), and the gray to the compressed octahedra (Ni$_{L}$). On the right the spin density of NNO. (b) On the top, the LSMO bulk structure is shown. On the bottom there is the spin density of the system. Images produced by VESTA software  \cite{VESTA}.
\label{fig:NNObulk}
}
\end{figure}

In Fig.~\ref{fig:NNObulk} we present the crystal structures of bulk LSMO and NNO and the correspondent spin mangetization obtained within our methodology, which is used for comparisons with the results obtained for the superlattices in the main text.

\nocite{*}

\bibliography{main}% Produces the bibliography via BibTeX.

\end{document}